\documentstyle[aps]{revtex}
\draft
\begin{document}
\title{Non-Newtonian effects in the peristaltic flow of a Maxwell fluid}
\author{David Tsiklauri$^1$ and Igor Beresnev$^2$}
\address{$^1$Space and Astrophysics Group, Physics Department, 
University of Warwick, Coventry, CV4 7AL, UK email: 
tsikd@astro.warwick.ac.uk; $^2$Department of Geological 
and Atmospheric Sciences,
Iowa State University, 253 Science I, Ames, IA 50011-3212,
U.S.A. email: beresnev@iastate.edu}
\maketitle
\begin{abstract}
We analyzed the effect of viscoelasticity 
on the dynamics of fluids in porous
media by studying the flow of a Maxwell fluid in a circular tube, 
in which the flow is induced by a wave traveling on the tube wall.
The present study investigates novelties
brought about into the classic peristaltic mechanism
by inclusion of non-Newtonian effects
that are important, for example, for hydrocarbons.
This problem has numerous applications in various branches of science,
including stimulation of fluid flow in porous media under the
effect of elastic waves.
We have found that in the extreme non-Newtonian regime
there is a possibility of a fluid flow in the direction {\it opposite}
to the propagation of the wave traveling on the tube wall.
\end{abstract}
\date{\today}
\pacs{47.55.Mh; 47.60.+i;  68.45.-v; 68.45.Kg; 92.10.Cg}

\section{Introduction}

Investigation of flow dynamics of a fluid in a tube having
circular cross-section, induced by a wave traveling  on its
wall (boundary), has many applications in various branches of
science. The physical mechanism of the flow induced by the traveling
wave
can be well understood and is known as the so-called
peristaltic transport mechanism. This mechanism is a natural cause
of motion of fluids in the body of living creatures, and it
frequently occurs in the organs such as ureters, intestines and
arterioles.
Peristaltic pumping is also used in medical instruments such as
heart-lung machine etc.[1].

Laboratory experiments have shown that an external sonic radiation
can considerably increase the flow rate of a liquid through a
porous medium (Refs.[1,2] and references therein).
Initially, the idea of flow stimulation via waves traveling on
the flow boundary, in the context of porous media, has been
proposed by Ganiev and collaborators [3].  
They proposed that sonic radiation generates traveling
waves on the pore walls in a porous medium.
These waves, in turn, generate net flow of fluid
via the peristaltic mechanism. Later,
this problem has been studied in a number of
publications, where authors used 
different simplifying assumptions in order to solve the
problem (see e.g. Ref.[4]).
The most recent and general study of stimulation of
fluid flow in porous media via peristaltic mechanism
is presented in Ref.[1], which we will use as a starting point
in order to include non-Newtonian effects into the peristaltic
model.

It is clear that  a usual peristaltic mechanism
discussed, e.g., in Ref.[1]
can be used to describe the 
behavior of a classic Newtonian fluid;
however, for example, oil and other hydrocarbons exhibit 
significant non-Newtonian behavior [5].
The aim of this paper is therefore to incorporate 
non-Newtonian effects
into the classical peristaltic mechanism [1].
Thus, the present work formulates a realistic 
model of the peristaltic mechanism which is 
applicable to the non-Newtonian fluids (e.g. hydrocarbons)
and not only to the Newtonian ones (e.g. ordinary water)
which have been extensively investigated in the past [1].

It should be noted that there were similar studies in the past
([6] and references therein). However, the previous contributions
discussed peristaltic mechanism for rheological equations other
than the Maxwellian one. Thus, the present study fills this gap 
in the literature. In addition, this study is 
motivated by the 
recent results of del Rio, de Haro and Whitaker [7]
and Tsiklauri and Beresnev [8], who found novel effects, including
the enhancement of a {\it Maxwellian fluid} flow in a tube that is
subjected to an oscillatory pressure gradient.

\section{The model}

We consider an axisymmetric cylindrical tube (pore) of radius $R$
and length
$L$. We assume that elastic wave induces a traveling wave on the
wall (boundary) of the tube 
with the displacement of the following form:
$$
W(z,t)=R+ a \cos( {{2 \pi}\over{\lambda}}(z-ct)), \eqno(1)
$$
where $a$ is the amplitude of the traveling wave, while
$\lambda$ and $c$ are its wave-length and velocity, respectively.
We note that $z$-axis of the ($r$,$\phi$,$z$) cylindrical
coordinate
system  is directed along the axis of the tube.

The equations which govern the flow are the balance of mass
$$
{{\partial \rho}\over{\partial t}}+ \nabla \cdot (\rho \vec v)=0,
\eqno(2)
$$
and the momentum equation
$$
\rho {{\partial \vec v}\over{\partial t}}+ \rho(\vec v \nabla)\vec
v=
- \nabla p - \nabla \tilde \tau,
\eqno(3)
$$
where $\rho$, $p$ and $\vec v$ are the fluid density, pressure and
velocity, respectively; $\tilde \tau$ represents the viscous stress
tensor.
We describe the viscoelastic properties of the fluid using the Maxwell's
model [7], which assumes that
$$
t_m {{\partial \tilde \tau}\over{\partial t}}= -\mu \nabla \vec  v
-{{\mu}\over{3}} \nabla \cdot \vec v - \tilde \tau, \eqno(4)
$$
where $\mu$ is the viscosity coefficient and $t_m$ is the
relaxation time.

We further assume that the following equation of state holds
$$
{{1}\over{\rho}}{{d \rho}\over{d p}}= \kappa, \eqno(5)
$$
where $\kappa$ is the compressibility of the fluid.
We also assume that the fluid's velocity has only $r$ and $z$
components. 

We make use of "no-slip" boundary
condition at the boundary of the tube, i.e.
$$
v_r(W,z,t)={{\partial W}\over{\partial t}}, \;\;\;
v_z(W,z,t)=0. \eqno(6)
$$

Eq.(4) can be re-written in the following form:
$$
\left(1+ t_m {{\partial  }\over{\partial t}}
\right) \tilde \tau= -\mu \nabla \vec  v 
-{{\mu}\over{3}} \nabla \cdot \vec v. \eqno(7)
$$
Further, we apply the operator $(1+ t_m {{\partial  }/ {\partial t}})$
to the momentum equation (3) and eliminate $\tilde \tau$ in it
using Eq.(7):
$$
-\left(1+ t_m {{\partial  }\over{\partial t}}
\right) \nabla p 
+ \mu \nabla^2 \vec v + {{\mu}\over{3}} \nabla(\nabla \cdot \vec v)
= \left(1+ t_m {{\partial  }\over{\partial t}}
\right) \left[ \rho {{\partial \vec v}\over{\partial t}}+ 
\rho(\vec v \nabla)\vec
v\right]. \eqno(8)
$$

The equations are made dimensionless by scaling the length by
$R$ and time by $R/c$. Also, we have introduced the following 
dimensionless variables (and have omitted the
tilde sign in the latter equations):
$\tilde W = W/R$, $\tilde \rho =\rho / \rho_0$, $\tilde
v_r=v_r/c$,
$\tilde v_z=v_z/c$, $\tilde p=p/(\rho_0c^2)$.
Here, $\rho_0$ is the regular (constant)
density at the reference pressure $p_0$.
We have also introduced $\epsilon = a/R$, $\alpha= 2 \pi R /
\lambda$,
$Re= \rho_0 c R/ \mu$, $ \chi = \kappa
\rho_0 c^2$.

Following Ref.[1], we seek the solution of the governing
equations in a form:
$$
p=p_0+\epsilon p_1(r,z,t) + \epsilon^2 p_2(r,z,t) + ...,
$$
$$
v_r=\epsilon u_1(r,z,t) + \epsilon^2 u_2(r,z,t) + ...,
$$
$$
v_z=\epsilon v_1(r,z,t) + \epsilon^2 v_2(r,z,t) + ...,
$$
$$
\rho= 1+ \epsilon \rho_1(r,z,t) + \epsilon^2 \rho_2(r,z,t) + ... \, .
$$

Then, doing a usual perturbative analysis using the latter
expansions, we can obtain a closed set of governing equations 
for the first ($\epsilon$) and second ($\epsilon^2$) order.

Further, following the authors of [1,8], we seek the 
solution of the liner problem in the form:
$$
u_1(r,z,t)=U_1(r)e^{i \alpha (z-t)}+ \bar U_1(r)e^{-i \alpha (z-t)},
$$
$$
v_1(r,z,t)=V_1(r)e^{i \alpha (z-t)}+ \bar V_1(r)e^{-i \alpha (z-t)},
$$
$$
p_1(r,z,t)=P_1(r)e^{i \alpha (z-t)}+ \bar P_1(r)e^{-i \alpha (z-t)},
$$
$$
\rho_1(r,z,t)=\chi P_1(r)e^{i \alpha (z-t)}+ \chi \bar P_1(r)e^{-i
\alpha (z-t)}.
$$
Here and in the following equations the bar denotes  a complex
conjugate.

On the other hand, we seek the second 
($\epsilon^2$) order solution in the form:
$$
u_2(r,z,t)=U_{20}(r)+U_2(r)e^{i 2\alpha (z-t)}+ \bar U_2(r)e^{-i 2
\alpha (z-t)},
$$
$$
v_2(r,z,t)=V_{20}(r)+V_2(r)e^{i 2\alpha (z-t)}+ \bar V_2(r)e^{-i 2
\alpha (z-t)},
$$
$$
p_2(r,z,t)=P_{20}(r)+P_2(r)e^{i 2\alpha (z-t)}+ \bar P_2(r)e^{-i 2
\alpha (z-t)},
$$
$$
\rho_2(r,z,t)=D_{20}(r)+D_2(r)e^{i 2\alpha (z-t)}+ \bar D_2(r)e^{-i 2
\alpha (z-t)}.
$$

The latter choice of solution is motivated by the fact that
the peristaltic flow is essentially a non-linear (second order)
effect [1], and adding a non-oscillatory term in the first order
gives only trivial solution. Thus, we can add non-oscillatory
terms, such as $U_{20}(r), V_{20}(r), P_{20}(r), D_{20}(r)$,
which do not cancel out in the solution after the time 
averaging over the period, only in the second and higher orders. 

In the first order by $\epsilon$ we obtain:
$$
-(1-i\alpha t_m)P_1' +{{1}\over{Re}}\left( 
U_1'' +{{U_1'}\over{r}} - {{U_1}\over{r^2}}
- \alpha^2 U_1\right) +
{{1}\over{3Re}}{{d}\over{dr}}
\left( U_1' + {{U_1}\over{r}} + i \alpha V_1 \right)=
-i \alpha (1-i \alpha t_m) U_1, \eqno(9)
$$
$$
-i \alpha (1-i\alpha t_m)P_1' +{{1}\over{Re}}\left( 
V_1'' +{{V_1'}\over{r}} 
- \alpha^2 V_1\right) +
{{i \alpha}\over{3Re}}
\left( U_1' + {{U_1}\over{r}} + i \alpha V_1 \right)=
-i \alpha (1-i \alpha t_m) V_1,\eqno(10)
$$
$$
\left( U_1' + {{U_1}\over{r}} + i \alpha V_1 \right)=
i \alpha \chi P_1. \eqno(11)
$$
Here, the prime denotes a derivative with respect to $r$.

Further, we re-write the system (9)-(11) in the following
form:
$$
-\gamma P_1' + \left( 
U_1'' +{{U_1'}\over{r}} - {{U_1}\over{r^2}}
- \beta^2 U_1\right)=0, \eqno(12)
$$
$$
-\gamma P_1 - {{i}\over{\alpha}}\left( 
V_1'' +{{V_1'}\over{r}} 
- \beta^2 V_1\right)=0, \eqno(13)
$$
where
$$
\gamma=(1-i \alpha t_m)Re - i \alpha \chi/3, \;\;\;
\beta^2=\alpha^2-i \alpha (1-i \alpha t_m) Re. \eqno(14)
$$
Note, that Eqs.(12)-(13) are similar to  Eq.(3.11) from
Ref.[1], except that $\gamma$ and $\beta$ are modified
by substitution $Re \to (1-i \alpha t_m)Re$.

Repeating the analysis similar to the one from Ref.[1],
we obtain  the master equation for $U_1(r)$ and find
its general solution as
$$
U_1(r)=C_1 I_1(\nu r) + C_2 I_1(\beta r), \eqno(15)
$$
where $I_1$ is the modified Bessel function of the first
kind of order 1; and $C_1$ and $C_2$ are complex constants
defined by
$$
C_1={{\alpha \beta \nu i I_0(\beta)}\over{2[\alpha^2I_0(\nu)I_1(\beta)-
\beta \nu I_0(\beta)I_1(\nu)]}}, \;\;\;
C_2={{-\alpha^3 i I_0(\nu)}\over{2[\alpha^2I_0(\nu)I_1(\beta)-
\beta \nu I_0(\beta)I_1(\nu)]}},
$$
where 
$$
\nu^2=\alpha^2{{(1-\chi)(1-i \alpha t_m)Re-(4/3)i \alpha \chi}
\over{(1-i \alpha t_m)Re-(4/3)i \alpha \chi}}.
$$
Here, $I_0$ is the modified Bessel function of the first
kind of order 0.

We also obtain the general solution for $V_1(r)$:
$$
V_1(r)={{i \alpha C_1}\over{\nu}}I_0(\nu r) +
{{i \beta C_2}\over{\alpha}}I_0(\beta r).
$$

The second-order solution $V_{20}(r)$ can also be found
in a way similar to the one used in Ref.[1]:
$$
V_{20}(r)=D_2-(1-i \alpha t_m)Re\int_{r}^{1}
[V_1(\zeta)\bar U_1(\zeta)+ \bar V_1(\zeta)U_1(\zeta)]d \zeta,
$$
where $D_2$ is a constant defined by
$$
D_2= - {{ i \alpha C_1}\over{2}}I_1(\nu)-
{{i \beta^2C_2}\over{2 \alpha}}I_1(\beta)
+{{-i \alpha \bar C_1}\over{2}}I_1(\bar \nu)+
{{i \bar{(\beta^2C_2)}}\over{2 \alpha}}I_1(\bar \beta).
$$

The net dimensionless fluid flow rate $Q$ can be calculated as [1]:
$$
Q(z,t)=2 \pi \left[ \epsilon \int_0^1 v_1(r,z,t) r dr +
\epsilon^2 \int_0^1 v_2(r,z,t) r dr + O(\epsilon^3)\right].
$$

In order to obtain the net flow rate averaged over one period of time,
we have to calculate
$$
<Q>= {{\alpha}\over{2 \pi}} \int_0^{2 \pi / \alpha} Q(z,t) dt.
$$
This time averaging yields
$$
<Q>= 2 \pi \epsilon^2 \int_0^1 V_{20}(r)r dr
$$
or finally substituting the explicit form of $V_{20}(r)$
we obtain for the dimensionless net flow rate
$$
<Q>=\pi \epsilon^2 \left[ D_2 - (1-i \alpha t_m)Re
\int_{0}^{1} r^2
[V_1(r)\bar U_1(r)+ \bar V_1(r)U_1(r)]d r \right]
\eqno(16)
$$

\section{Numerical Results}

In the previous section, we have shown that the inclusion of
non-Newtonian effects into the classical peristaltic mechanism
by using the Maxwell fluid  model yields the following change: 
$Re \to (1-i \alpha t_m)Re$ in all of the solutions.

It is known that the viscoelastic fluids described by the
Maxwell model have different flow regimes depending on the value
of the parameter $De=t_v/t_m$, which is called the Deborah number [7].
In effect, Deborah number is a ratio of the characteristic time
of viscous effects $t_v=\rho R^2/ \mu$ to the relaxation time
$t_m$. As noted in Ref. [7], the value of the parameter $De$
(which the authors of Ref.[7] actually call $\alpha$)
determines in which regime the system resides. Beyond 
a certain critical value ($De_c=11.64$), the system is
dissipative, and conventional 
viscous effects dominate. On the other hand, for
small $De$ ($De < De_c$) the system
exhibits viscoelastic behavior. 

A numerical code has been written to calculate $<Q>$
according to  Eq.(16).
In order to check the validity of our code, we run it for the parameters
similar to the ones used by other authors.
For instance, for $\epsilon=0.15$, $Re=100.00$, $\alpha=0.20$,
$\chi=0$, $t_m=0$ we obtain $<Q>=0.2708706458$, which
is equal (if we keep 4 digits after the decimal point) to the result 
of the authors of Ref.[1] who actually obtained $<Q>=0.2709$.

Further, we have made several runs of our code for
different values of the parameter $t_m$.
We note again that $t_m$ enters the equations
because we have included non-Newtonian effects into
our model. Eq.(16) will be identical to the
similar Eq.(4.1) from Ref.[1] if we set $t_m=0$ in all our
equations.

The results of our calculation are presented in Fig.1,
where we investigate the dependence of $<Q>$ on the
compressibility parameter $\chi$ for the various
values of $t_m$.
In order to compare our results with the ones from Ref.[1],
we have plotted $<Q>$ for the following set of parameters:
$\epsilon=0.001$, $Re=10000.00$, $\alpha=0.001$,
$t_m=0$ (solid line). We note that the curve is identical
to the corresponding curve in Fig.2 from Ref.[1].
This obviously corroborates the validity of our numerical code.
Further, to investigate the dependence of the flow rate $<Q>$
on $t_m$, we perform the calculation for a few
values of $t_m$. When $t_m=1.0$, we notice no noticeable
change in the plot as both curves coincide within the
plotting accuracy. For $t_m=100.00$ (dashed curve with
crosses), we notice slight deviation from the Newtonian 
limiting case (solid line),
which translates into shifting the maximum towards
larger $\chi$'s. For $t_m=1000.00$ (dash-dotted curve with
asterisks), we notice further deviation from the Newtonian flow,
which also translates into shifting the maximum towards
larger $\chi$'s. However, for $t_m=10000.00$ (dashed
curve with empty squares), we note
much more drastic changes, including the absence of a maximum
and rapid growth of $<Q>$ in the considered interval
of variation of compressibility parameter $\chi$. The
observed pattern conforms to our expectation,
since large $t_m$ means small $De$ ($De < De_c$) and 
the system exhibits strong viscoelastic behavior.
Note that $t_m$ is dimensionless and scaled by $R/c$.

After the above discussion it is relevant to define
quantitatively the transition point where the flow
starts to exhibit (non-Newtonian) viscoelastic effects.
It is known [7] that $De=t_v/t_m=(\rho R^2)/(\mu t_m)$.
Now, using definition of $Re= \rho c R / \mu$ we
can define critical value of $t_m$ as 
$$
t_{mC}= \left({{Re}\over{De_c}} \right)  {{R}\over{c}}.
\eqno(17)
$$
In all our figures we have used $Re=10000.0$.
If we put the latter value of $Re$ and the critical value
of the Deborah number 11.64 [7] into Eq.(17) we obtain 
$t_{mC}= 859.11$ (measured in units of $R/c$).
Therefore, the values of $t_m$ greater than $t_{mC}$ 
(for a given $Re$) correspond to sub-critical ($De < De_c$)
Deborah numbers for which viscoelastic effects
are pronounced.

In Fig.2 we investigate the behavior of the flow rate
$<Q>$ on the parameter $\alpha$, which is the tube radius measured
in wave-lengths. Again, to check for the consistency of
our numerical results with the ones from Ref.[1] 
and also investigate novelties brought
about by the introduction of non-Newtonian effects 
(appearance of non-zero relaxation time $t_m$) into
the model, we first plot $<Q>$ versus $\alpha$ for the
following set of parameters:
$\epsilon=0.001$, $Re=10000.00$, $\chi=0.6$,
$t_m=0$. If we compare the solid curve in our Fig.2 
with the dashed curve in Fig.3 of Ref.[1], we will 
note no difference, which again corroborates the
validity of our numerical code. 
We then set $t_m$ to various non-zero values and 
investigate the changes introduced by non-Newtonian effects.
As in Fig.1, we notice no change for $t_m=1.0$.
For $t_m=100.00$ and $t_m=1000.00$, we notice 
that flow rate somewhat changes attaining lower
values as $\alpha$ (radius of the tube) increases.

We treat the latter, $t_m=1000.00$, case separately for
the reason of an appearance of a novel effect of
{\it negative} flow rates when the interval of 
variation of $\alpha$ is increased up to 0.05. 
Again,  we expect that for 
large $t_m$ ($t_m > t_{mC}=859.11$), i.e. small $De$ ($De < De_c$), 
the system should exhibit viscoelastic behavior.
We note from Fig.3 that for $\alpha \geq 0.035$,
$<Q>$ becomes {\it negative}, i.e., we observe backflow.
In fact, by doing parametric study we conclude that 
$t_{mC}$ is the critical value of $t_m$
above which we observe backflow.
By increasing $t_m$ further, $t_m=10000.00$,
we note from Fig.4 that in this deeply non-Newtonian
regime $<Q>$ becomes highly oscillatory, but what is
unusual again is that we observe the  negative flow rates
for certain values of $\alpha$ that is the tube radius measured
in wave-lengths.
Obviously, the negative $<Q>$ means that flow occurs in
the direction opposite to the direction of 
propagation of traveling wave on the tube
wall. Oscillatory behavior (appearance of numerous
of maxima in the behavior of a physical variable)
in the deeply non-Newtonian regime is not new [8].
However, the flow of a fluid created by the peristaltic
mechanism in the direction opposite to the direction of 
propagation of traveling wave
is unusual and should be attributed to a
complicated, non-linear form of the response of a Maxwell 
fluid to the stress exerted by the wave.

\section{Conclusions}

In this paper, we investigated the dynamics of fluid flow
in a tube with a circular cross-section, induced by a wave
traveling on its wall (boundary). This problem has numerous 
applications in various branches of science,
including stimulation of fluid flow in porous media under the
effect of elastic waves.

The present study investigates novelties
brought about into the classic peristaltic mechanism
by the inclusion of non-Newtonian effects based on the
model of Maxwell fluid.

We have found that in the extreme non-Newtonian regime
there is a possibility of flow in the direction opposite
to the propagation of the wave traveling on the tube wall.
Somewhat similar effect is known as the acoustic streaming [9],
in which an acoustic wave propagating in a tube induces
a mean flow in the direction of propagation in the
acoustic boundary layer, but in the
opposite direction in the central part of the tube.
The mean flow or acoustic streaming is caused by the
presence of friction at the bounding surfaces of the
tube. While fluid away from the neighborhood of a boundary
vibrates irrotationally as the acoustic wave passes,
fluid in close proximity to the boundary must vibrate
rotationally to satisfy the no-slip condition on the
tube wall. This deviation from inviscid, irrotational 
behavior provides an effective driving force known
as the Reynolds stress. This effective force, because
of it is quadratic rather than linear, has non-vanishing
time-average tangential component to the tube wall
that drives flow in the boundary layer.
In the case considered in our paper instead of having
acoustic wave propagating  through the volume of the tube,
we have a wave traveling of the tube wall, besides 
we have further complication of considering non-Newtonian
(Maxwell) fluid (recall that the discovered effect is 
demonstrated for the case of  non-Newtonian
regime $t_m/t_{mC} > 1.0$). 
Similarly,
the peristaltic flow itself is a second order non-linear 
effect. Therefore, the flow of a fluid created by the peristaltic
mechanism in the direction opposite to the direction of 
propagation of traveling wave (i.e. back-flow) could be
explained by a
complicated, non-Newtonian, non-linear response of a Maxwell 
fluid to the stress exerted by the traveling wave.

\acknowledgments

This work was supported by the Iowa State University Center for
Advanced Technology Development and ETREMA Products, Inc.
We would like to thank anonymous referee for useful suggestions
of improvement of the manuscript.

\centerline{\bf Figure captions}

Fig. 1 Dimensionless flow rate $<Q>$ as a function of compressibility
parameter $\chi$. The parameters used
are $\epsilon=0.001$, $Re=10000.00$, $\alpha=0.001$.
$t_m=0$ corresponds to the solid line, whereas
$t_m=100.00, \; 1000.00$ and $10000.00$ correspond 
to the dashed curve with
crosses, dash-dotted curve with
asterisks, and dashed curve with empty squares, respectively.

Fig. 2 Plot of dimensionless flow rate $<Q>$ 
as a function of $\alpha$. Here, $\epsilon=0.001$, 
$Re=10000.00$, $\chi=0.6$.
$t_m=0$ corresponds to the solid line, whereas
$t_m=100.00$ and $1000.00$ correspond to the dashed curve with
crosses and dash-dotted curve with
asterisks, respectively.

Fig. 3 Plot of dimensionless flow rate $<Q>$ 
as a function of $\alpha$ on a larger than in Fig.2
interval of variation of $\alpha$. Here, $\epsilon=0.001$, 
$Re=10000.00$, $\chi=0.6$, $t_m=1000.00$.

Fig. 4 Plot of dimensionless flow rate $<Q>$ 
as a function of $\alpha$. Here, $\epsilon=0.001$, 
$Re=10000.00$, $\chi=0.6$, $t_m=10000.00$.
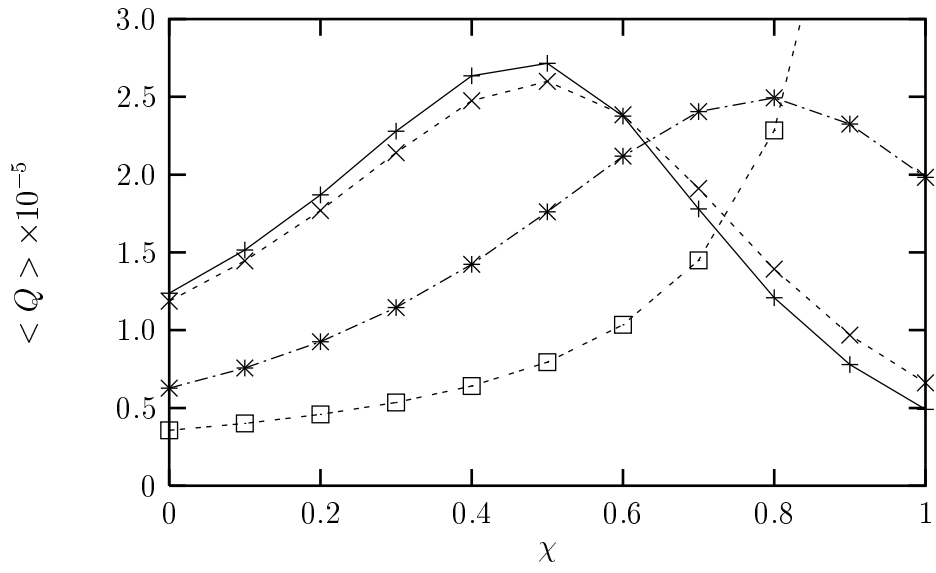
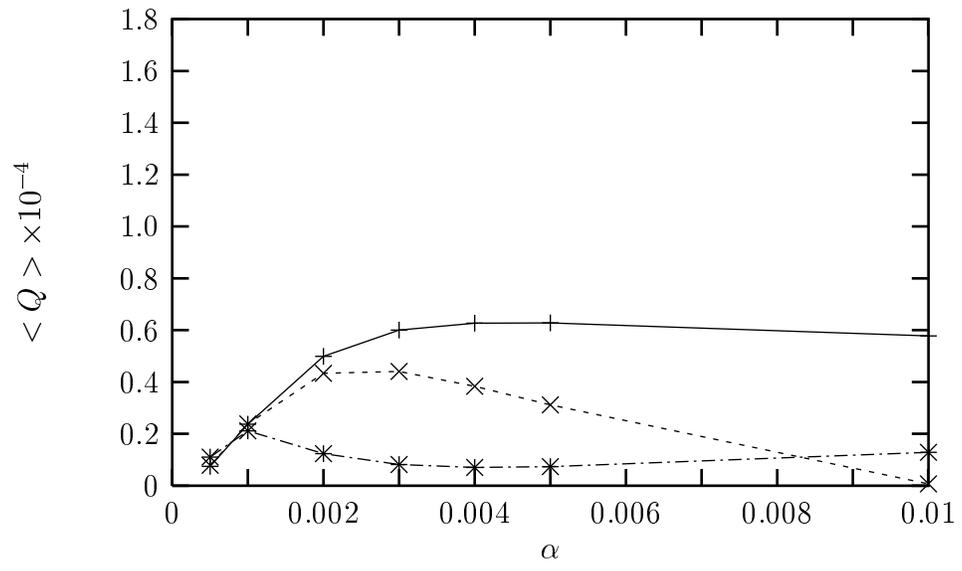
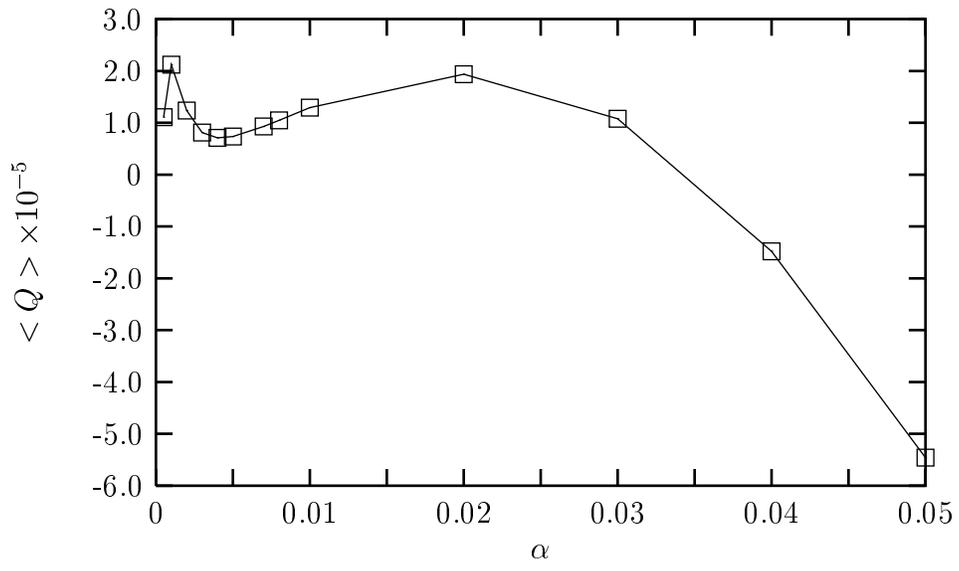
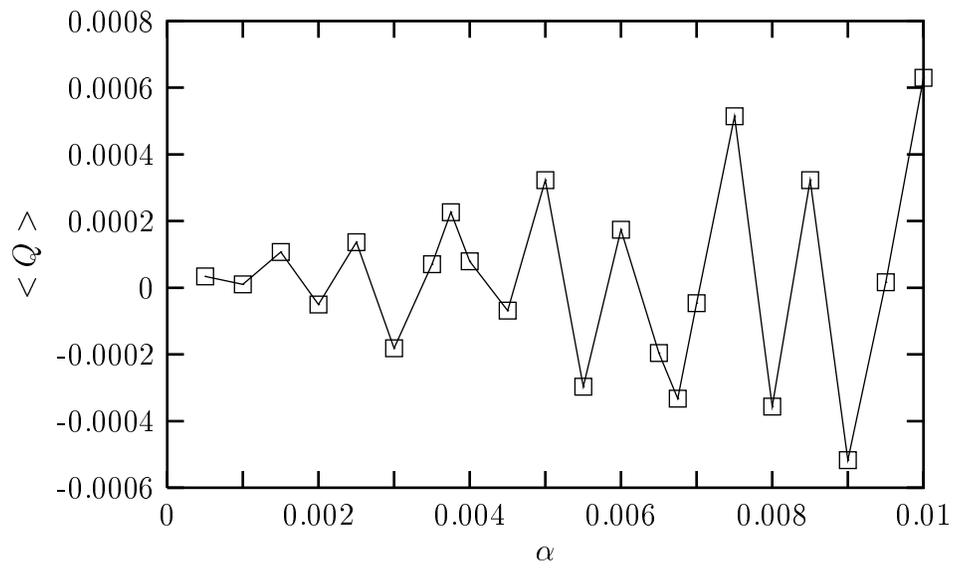
\end{document}